%% file: 010_main.tex
\documentclass[english,sigconf,authorversion]{acmart}

\renewcommand\footnotetextcopyrightpermission[1]{} 
\usepackage[T1]{fontenc}
\usepackage[latin9]{inputenc}
\PassOptionsToPackage{ruled}{algorithm2e}
\usepackage{color}
\usepackage{babel}
\usepackage{booktabs}
\usepackage{algorithm2e}
\usepackage{graphicx}
\PassOptionsToPackage{normalem}{ulem}
\usepackage{ulem}
\ifx\hypersetup\undefined
  \AtBeginDocument{%
    \hypersetup{unicode=true}
  }
\else
  \hypersetup{unicode=true}
\fi

\makeatletter

\providecommand{\tabularnewline}{\\}
\newcommand{\lyxdot}{.}

\newcommand{\redefineshortauthors}[1]{\renewcommand{\shortauthors}{#1}}

\SetAlFnt{\small}
\SetAlCapFnt{\small}
\SetAlCapNameFnt{\small}
\SetAlCapHSkip{0pt}
\IncMargin{-\parindent}

\copyrightyear{2019}
\acmYear{2019}
\setcopyright{none}

\makeatother

\begin{document}
\title[DSL-based Code Suggestions for Data Analysis]{One DSL to Rule Them All: \\
IDE-Assisted Code Generation for Agile Data Analysis}
\author{Artur Andrzejak}
\orcid{0000-0003-0150-8220}
\affiliation{\institution{Heidelberg University}\city{Heidelberg}\country{Germany}}
\email{artur@uni-hd.de}
\author{Oliver Wenz}
\orcid{todo}
\affiliation{\institution{Heidelberg University}\city{Heidelberg}\country{Germany}}
\email{o.wenz@stud.uni-heidelberg.de}
\author{Diego Costa}
\orcid{todo}
\affiliation{\institution{Heidelberg University}\city{Heidelberg}\country{Germany}}
\email{diego.costa@informatik.uni-heidelberg.de}
\begin{abstract}
\input{011-abstract.tex}

\end{abstract}




\thanks{}

\maketitle

\redefineshortauthors{A. Andrzejak et al.}

\section{Introduction} \label{sec:Introduction}

Data analysis studies involve many steps of data manipulation and
analysis. Typically, software tools with fixed processing workflows
are not flexible enough to cover a large variety of scenarios encountered
in research, and so some form of programming is mandatory. Consequently,
users frequently create project-specific analysis pipelines using
various frameworks and tools. These range from data-flow programming
environments like KNIME to interactive data processing frameworks
like JupyterLab (with Pandas, Scikit-Learn, TensorFlow etc.) to frameworks
capable of massively parallel data processing like Apache Spark.

Most approaches are characterized by an inherent trade-off between
ease-of-use and flexibility and/or performance. For example, due to
the low development effort, Matlab is popular in industry and science
to prototype data-centric applications. However, for the production
scenario or to process massive data sets, such prototypes need to
be rewritten from scratch into C/C++ or distributed computing frameworks
like Apache Spark. In general, researchers and practitioners are still
faced by multiple challenges at the intersection of data science and
software engineering:

\emph{Programming barrier}. Proprietary data formats and specific
requirements can imply a substantial amount of project-specific script
programming.  This seriously increases the duration and cost of data
analysis projects, and makes it substantially harder for domain specialists
(frequently with only limited programming skills) to interact with
the data directly.

\emph{Reuse problem}. A majority of the above-mentioned programming
effort is put into adjusting code to the specifics of the project
and its data sets. New and more widely applicable algorithms and libraries
are frequently result of an additional coding effort performed only
as a \textquotedblleft side-effect\textquotedblright{} of a project.
This creates an unfavorable overall ratio between reusable (library-like)
and project-specific code bases.

\emph{Scalability problem.} Implementing sequential versions of algorithms
and software pipelines can be already quite challenging, yet their
scalable (massively parallel) versions for large data sets are typically
significantly more complex. Moreover, scalable versions require other
data structures, libraries, and even programming paradigms (such as
DAG operation graphs in Apache Spark), and so significant costs are
needed (typically complete re-implementation) to provide scalable
versions of software pipelines. Therefore, users frequently work with
sequential scripts even if parallel processing would be beneficial,
which can limit the considered data sizes and slow down the execution.

Multiple efforts have been undertaken to address these challenges,
including novel human-in-the-loop approaches like Predictive Interaction
\citep{heer2015predictive}, development of new programming languages
like Julia \citep{bezanson2015abstraction}, and research on advanced
methods like program synthesis for data analysis \citep{gulwani2016programming},
\citep{barowy2015flashrelate}, \citep{jin2017foofahtransforming}.
Nevertheless, many of these techniques address only selected special
cases, are still in the research phase, or might require to migrate
to a new platform or a programming environments.   To our knowledge,
for a majority of application scenarios there are still no comprehensive
and effective solutions, or solutions which are compatible with \textquotedbl legacy\textquotedbl{}
software (i.e. frameworks, libraries, and programming languages),
which would complement or support existing software ecosystems instead
of trying to replace them.

We propose an approach which  supports the development of data science
scripts during the editing process. In essence, our approach can be
understood as advanced code recommendations, where users formulate
their intention in an abstract and user-friendly Domain Specific Language
(DSL) for dataframe (table) manipulation and analysis. Such DSL statements
are directly translated into executable Python code while DSL can
remain in the scripts as comments.

Two elements of our solution are essential in addressing the above-mentioned
challenges. First, we provide intelligent editing support for the
DSL (code completion). Together with a self-explaining design of our
DSL this substantially lowers the initial effort to command the DSL
and reduces the adoption barrier. Furthermore, the very same DSL can
generate - according to user settings - code for various frameworks
(``targets''). Our prototype currently offers code generation for
Pandas (a popular Python library for in-memory processing and analysis
of time series and tables), and for Apache PySpark (Python bindings
for Spark, a framework for distributed processing of massive data
sets). Generating code for multiple targets supports converting of
Pandas scripts to Spark in scenarios where users first experiment
and prototype on small data sets (using Pandas) but later need a scalable
solution based on Spark (or vice-versa).

DSLs are becoming increasingly popular in software solutions \citep{fowler2010domainspecific},
\citep{kats2010thespoofax}, \citep{campagne2016themps}, \citep{bettini2016implementing},
\citep{dejanovic2017textxa}, as they raise the abstraction level
of code and facilitate communication with domain experts. So-called
\emph{external} DSLs are completely independent languages, with code
typically not mixed with other languages. While such DSLs allow high
syntactic flexibility, adjustment to the target domain, and IDE support,
the effort of implementing functionality beyond the original DSL intention
can be substantial. Typically, functionality extensions are only possible
via custom User Defined Functions (UDFs) or by embedding the DSL into
a general-purpose language via strings or separate files (similarly
to how SQL is used from C++ or Java). In such cases, developers must
learn and use additional APIs to interact between two languages. In
addition, debugging as well as (static) code analysis become cumbersome.

Another approach is to use \emph{internal} DSLs which are essentially
libraries written in general-purpose programming languages with flexible
syntax, e.g. Ruby, Scala, Kotlin, or F\#. Such DSLs are easier to
implement and avoid the interoperability issues, but have only a constrained
syntax and no IDE support (Intellisense etc.).  In all cases, using
a DSL can lead to a lock-in effect and might be a barrier to new developers
on a project.

In our solution we use an external DSL yet attempt to circumvent the
above-mentioned issues for such languages by allowing a ``no-barrier''
coexistence of the DSL and the generated code. First, developers can
complement and modify the generated code directly during editing,
which prevents the problem of lacking DSL functionality. Our DSL is
intended to provide support only for common, frequently used operations,
while more specific requirements are implement in the target language.
Furthermore, developers are free to decide on how to use our DSL:
it can serve only as ``active help'' during editing; it can be used
as comments/explanations for the generated code; or it can serve as
primary source code for simpler scenarios. We believe that this flexibility
can be helpful in the adoption of our solution and reduce the lock-in
effect. An obvious disadvantage of our proposal is that the DSL code
and the generated code might become out-of-sync during editing, in
worst case leading to a misleading code description (similarly to
an outdated documentation). We will address this issue in further
research, e.g. by automatically flagging DSL code which no longer
fits to the generated target code.

This paper is organized as follows. Section~\ref{sec:Approach} describes
the details of the approach. Section~\ref{sec:Implementation} outlines
the implementation. We present the preliminary evaluation in Section~\ref{sec:Evaluation}
and discuss related work in Section~\ref{sec:Related-Work}. Section~\ref{sec:Future-Work-and-Conclusions}
describes conclusions and future work.

\section{Approach} \label{sec:Approach}

\begin{figure}
\includegraphics[clip,width=1\linewidth]{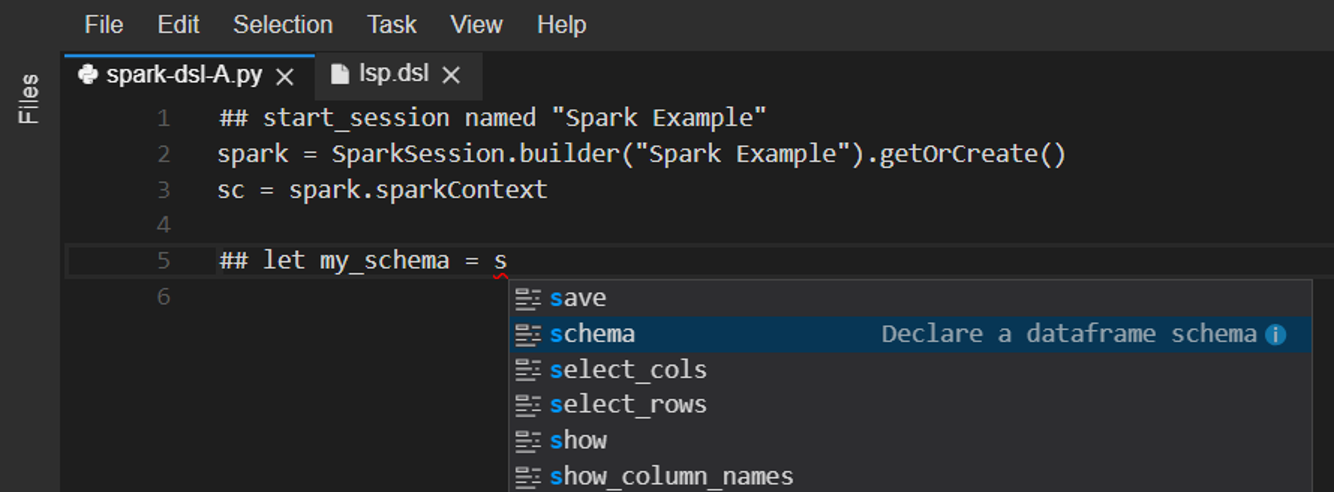}\caption{\label{fig:interaction-A}Screenshot of an exemplary editing session.}
\end{figure}
Our approach supports script developers via an IDE-supported DSL for
generating Python code for common data science tasks. Users enter
DSL statements as pseudo-comments while being assisted by code completions
(Figure~\ref{fig:interaction-A}), and can choose to insert executable
code generated from these DSL statements. It is also possible to specify
the type of generated code, or the type of the target framework (currently
Pandas or PySpark). Apart from the changed editing experience (noticeable
only in lines starting with the DSL prefix code, here \texttt{\#\#}),
there is no difference to a normal development and execution process.
By placing DSL code in comments, normal execution and testing flow
is not influenced and does not need to be changed. As noted in the
introduction, users can deploy our tool in any editor/IDE supporting
the Language Server Protocol, which currently covers all major IDEs.
Consequently, we estimate the barrier to the adoption of the tool
to be low.

The utility of this approach is largely determined by the design of
the DSL, in particular, the power of the DSL (or its``expressiveness'')
in terms of common operations for the target frameworks. To ensure
a high level of DSL coverage, we have analyzed several popular ``cheat-sheets''
for Pandas and PySpark, as well as some tutorials for these frameworks.
Based on this analysis we designed a DSL which attempts to cover most
of the elements found in these sources, see Table~\ref{tab:DSL-syntax}.
It should be noted that we purposefully do not attempt to cover all
of the functionality in our DSL. This would largely increase the effort
of the implementation (DSL grammar design and code generation), and
would make the DSL fragile to changes in the targeted frameworks.
Instead, we assume that developers will implement more specific functions
directly in the Python code.

\subsection{DSL for dataframe operations} \label{subsec:Table-manipulation-DSL}

Our DSL attempts to be easy-to-understand (or, in the best case, even
self-explanatory) yet concise. Thanks to code completion, long keywords
are acceptable, and so we preferred better readability in the DSL
design than compact but ambiguous keywords. The hurdle of learning
and understanding the DSL is further reduced by explanations of the
commands provided in the list of suggestions (see Figure~\ref{fig:interaction-A}).

Table~\ref{tab:DSL-syntax} gives an overview of the essential parts
of our DSL. It covers functionality for data I/O, simple selection,
deleting rows or columns, aggregation, multiple types of data transformations,
and essential data inspection. In addition, we provide operations
which are suitable only for Spark code generation (in Pandas mode,
these emit empty code). The DSL has also a meta-command \textbf{target\_code}
\uline{spark} | \uline{pandas} which specifies for which framework
code should be generated.

Many modern data science libraries allow the \emph{Fluid API }coding
style, i.e. chaining of method calls like in \texttt{var.func\_A().func\_B()}
to avoid need for intermediate variables (in R, similar effect is
achieved by the ``pipeline'' operator '\texttt{\%>\%}'). Our DSL
also allows this style, with ``:'' chosen as the ``pipeline''
operator (this can be easily changed). Furthermore, to specify the
dataframe on which the operation chain is to be performed, we use
\textbf{on} \emph{dataframe} syntax. In Table~\ref{tab:DSL-syntax},
this is used in several examples (e.g. line 2).

\begin{table}
\caption{\label{tab:DSL-syntax}DSL overview. Keywords are in \textbf{bold},
choices from a list are \uline{underlined}, and other parameters
are \emph{emphasized}.}

\begin{tabular}{ll}
\toprule 
Category & DSL examples (prefix \textbf{\#\#} is omitted)\tabularnewline
\midrule 
I/O operations & \emph{result} = \textbf{load} \textbf{as} \uline{json} \emph{'some\_path.json}'\tabularnewline
 & \textbf{on} \emph{df} : \textbf{save} \textbf{as} \uline{csv} \textbf{to}
'\emph{some\_path.csv}'\tabularnewline
Selection & \emph{result} = \textbf{on} \emph{df} : \textbf{select\_cols} \emph{a},
\emph{b}, \emph{c}\tabularnewline
 & ...\textbf{ select\_rows} \emph{col1 == m }\textbf{or}\emph{ col2
< 3}\tabularnewline
 & ...\textbf{ select\_rows} \emph{col1 > 0 }\textbf{and}\emph{ col3
}\textbf{in}\emph{ }{[}\emph{v1, v2, v3}{]}\tabularnewline
Deletion & \emph{result} = \textbf{on} \emph{df} : \textbf{drop\_cols} \emph{x},
\emph{y}, \emph{z}\tabularnewline
 & \emph{...} \textbf{drop\_rows} \emph{col1 > 0 }\textbf{and }\emph{col2
}\textbf{not in}\emph{ }{[}\emph{v1, v2}{]}\tabularnewline
Aggregation & \emph{result} = \textbf{on} \emph{df} :\textbf{ group\_by} \emph{col1}\textbf{
apply}\emph{ }\uline{sum}\tabularnewline
 & ...\textbf{ group\_by} \emph{col1, col2 }\textbf{apply}\emph{ }\uline{min}\tabularnewline
Transform. & \emph{result} = \textbf{on} \emph{df} :\textbf{ on\_missing} \textbf{fill\_with}\emph{
value}\tabularnewline
 & ...\textbf{ on\_missing} \textbf{drop\_rows}\tabularnewline
 & ...\textbf{ replace }\emph{old\_value}\textbf{ by}\emph{ new\_value}\tabularnewline
 & ...\textbf{ apply\_fun }\emph{function}\textbf{ on}\emph{ }\uline{cols}\tabularnewline
 & ...\textbf{ apply\_fun }\emph{function}\textbf{ on}\emph{ }\uline{rows}\tabularnewline
 & ...\textbf{ append\_col }\emph{col\_name}\textbf{ }\tabularnewline
 & ...\textbf{ sort\_by }\emph{col\_name}\tabularnewline
 & ...\textbf{ drop\_duplicates}\tabularnewline
 & ... \textbf{rename\_cols} \emph{c1} \textbf{to} \emph{p}, \emph{c2}
\textbf{to} \emph{q}\tabularnewline
Inspection & \textbf{on} \emph{df} : \textbf{show}\tabularnewline
 & \textbf{on} \emph{df} : \textbf{describe}\tabularnewline
 & ... \textbf{return\_top\_N }\emph{10}\tabularnewline
 & ...\textbf{ select\_rows} \emph{col1 == m }\textbf{: count}\tabularnewline
Spark only & \textbf{start\_session} \textbf{named} 'session name'\tabularnewline
 & \textbf{stop\_session}\tabularnewline
 & \emph{s }= \textbf{schema} \emph{col1} \textbf{of} \uline{int},
\emph{col2} \textbf{of} \uline{str}\tabularnewline
 & \emph{result} = \textbf{load}\emph{ 'some\_path.txt' }\textbf{with\_schema}\emph{
s}\tabularnewline
Pandas only & ...\textbf{ append\_row }\emph{col\_name }\textbf{default }\emph{default\_val}\tabularnewline
Options & \textbf{target\_code }= \uline{spark}\tabularnewline
 & \textbf{target\_code }= \uline{pandas}\tabularnewline
\bottomrule
\end{tabular}
\end{table}

\subsection{Generated code} \label{subsec:approach-generated-target-code}

As explained in Section~\ref{subsec:Architecture}, DSL code is translated
into executable (Python) code by our tool when a user requests a code
completion action and the current DSL statement is syntactically correct.
We show in Table~\ref{tab:examples_gen_code} some examples of generated
code. Typically, the DSL code and the target code have similar structure,
e.g. order of parameters. Therefore, mapping of the DSL code to the
target code is relatively easy. For example, in most cases we can
map a DSL syntax node (or grammar rule) to the target code without
considering other syntax nodes.

\begin{table}
\caption{\label{tab:examples_gen_code}Examples of generated target code. In
the column \emph{Type}, 'P' indicates Pandas code, and 'S' Spark code.}

\begin{tabular}{ll}
\toprule 
Type & Code (for DSL, prefix \textbf{\#\#} is omitted)\tabularnewline
\midrule 
DSL & x = \textbf{load} \textbf{as} csv some\_path\tabularnewline
P & x = pd.read\_csv(some\_path)\tabularnewline
DSL & x = \textbf{load} \textbf{as} csv some\_path \textbf{with\_schema}
S\tabularnewline
S & x = spark.read.csv(some\_path, schema=S)\tabularnewline
\midrule 
DSL & x = \textbf{on} y : \textbf{select\_cols} a, b, c : \textbf{count}\tabularnewline
P & x = y{[}{[}'a', 'b', 'c'{]}{]}.count()\tabularnewline
S & x = y.select('a', 'b', 'c').count()\tabularnewline
\midrule 
DSL & x = \textbf{on} y : \textbf{select\_rows} col1 == m \textbf{and} col3
\textbf{in} {[}v1, v2, v3{]}\tabularnewline
P & x = y{[}(y.col1 == m) \& (y.col3.isin({[}v1, v2, v3{]})){]}\tabularnewline
S & x = y.filter((y.col1 == m) \& (y.col3.isin({[}v1, v2, v3{]})))\tabularnewline
\midrule 
DSL & x = \textbf{on} y :\textbf{ rename\_cols} c1 \textbf{to} p, c2 \textbf{to}
q\tabularnewline
P & x = y.rename(columns=\{'c1': 'p', 'c2': 'q'\})\tabularnewline
S & x = y.withColumnRenamed('c1', 'p').with...('c2', 'q')\tabularnewline
\bottomrule
\end{tabular}
\end{table}

\section{Implementation} \label{sec:Implementation}

\subsection{Architecture} \label{subsec:Architecture}

\begin{figure}
\includegraphics[viewport=0bp 405bp 590bp 540bp,clip,width=1\linewidth]{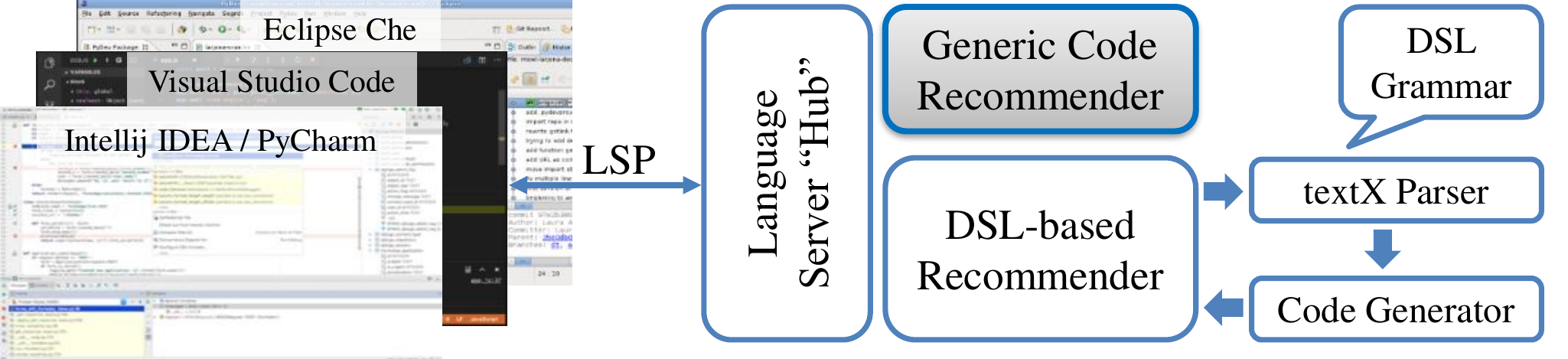}\caption{\label{fig:architecture}Architecture of our prototype.}
\end{figure}
Figure~\ref{fig:architecture} outlines the architecture of our prototypical
implementation. To reach developers with various preferred code editors
or IDEs, we use a Language Server Protocol (LSP) \citep{microsoftcorp.2019language}.
LSP decouples a particular editor/IDE from the ``coding services''
(including code completion, linting and basic refactoring) via a JSON-RPC
interface. Editors/IDEs supporting LSP (clients) include Visual Studio
(Code), Jetbrain products (PyCharm, Ryder, Intellij IDEA,...), Eclipse
(Che), Vim, Emacs, and others. Similarly, there exist a large number
of language servers for major programming languages.

Our tool emulates a Language Server via a thin layer (``Hub'') which
is responsible to dispatching client requests to sub-modules, and
forwarding responses from these sub-modules to the client. The dispatch
mechanism essentially analyzes a code completion request (in LSP,
\texttt{textDocument/completion} request) whether it has been issued
in a code line containing DSL-code, or not. In the DSL case, a request
is handled by our DSL-based Recommender (see Figure~\ref{fig:architecture}),
otherwise forwarded (unchanged) to a regular language server (for
Python, our prototype uses a server by Palantir Technologies\footnote{\href{https://github.com/palantir/python-language-server}{https://github.com/palantir/python-language-server}}).
Few other LSP requests are duplicated to all sub-modules to ensure
coherency of workspace/file caching, in particular \texttt{DidChangeTextDocument}.

The DSL-based Recommender uses textX \citep{dejanovic2017textxa}
and textX-LS\footnote{\href{https://github.com/textX/textX-LS}{https://github.com/textX/textX-LS}}
to implement code completion suggestions for our DSL grammar. textX
is a Python library for defining and implementing DSLs, and textX-ls
is a generic language server implementation which provides syntax
checking and code completion for any language defined with textX.
A response from textX-ls is a list of ranked recommendations as shown
in Figure~\ref{fig:interaction-A}. We add to this list a preview
of the generated target code (if DSL syntax is valid) with high rank,
i.e. at position 1. In this way, a user can insert the generated code
without breaking the editing flow.

\subsection{Code generation\label{subsec:Code-generation}}

\begin{figure*}
\includegraphics[viewport=38bp 40bp 915bp 298bp,clip,width=0.99\linewidth]{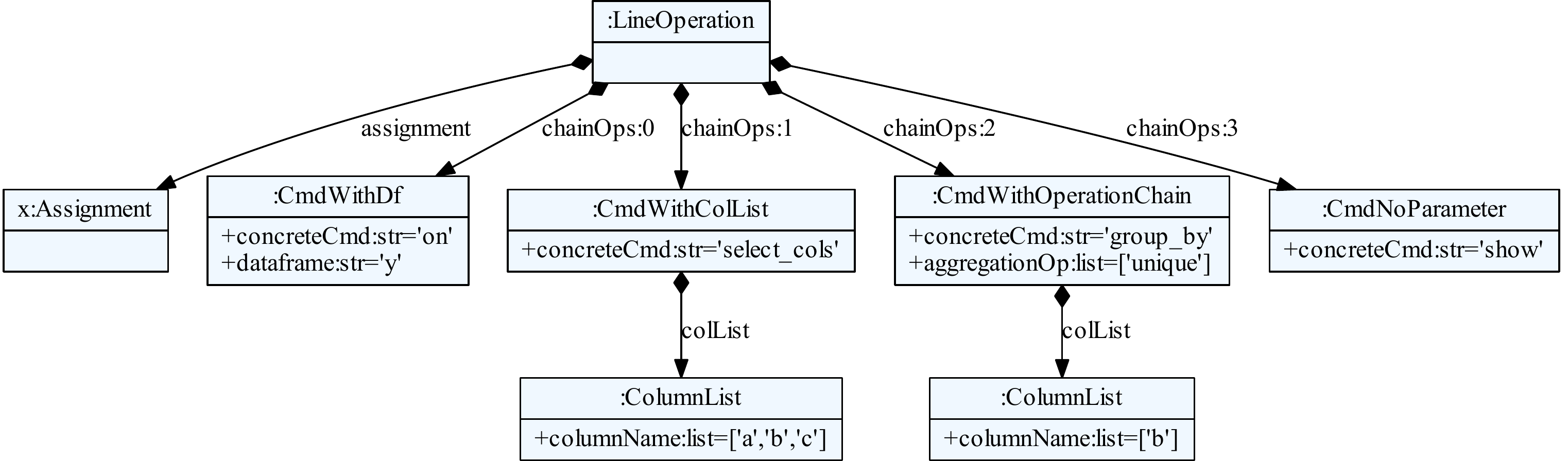}\caption{\label{fig:ast-example}textX-generated Python objects for the DSL
code ``x = \textbf{on} y : \textbf{select\_cols} a, b, c : \textbf{group\_by
}b\textbf{ apply }unique\textbf{ : show}''.}
\end{figure*}
Code generation is implemented as a set of few classes on top of the
textX library. Upon start of our tool, the DSL definition is parsed
and remains in memory as a meta-model (note that contrary to other
DSL construction tools like Xtext or MPS, grammar is parsed dynamically,
without generating code for a parser/lexer). For each LSP request
with a DSL in the current editor line we use textX to parse this line
with the prepared meta-model. Upon success, textX returns a tree of
Python objects corresponding to an Abstract Syntax Tree (AST).

To generate code, we essentially parse this tree using recursion,
iteration and if/else statements, and collect the generated code using
a string buffer. Figure~\ref{fig:ast-example} shows an example AST
for the DSL ``x = \textbf{on} y : \textbf{select\_cols} a, b, c :
\textbf{group\_by }b\textbf{ apply }\uline{unique}\textbf{ : show}''.
A Python method processes the AST-node \emph{LineOperation} by inspecting
a corresponding Python object \emph{obj} dynamically generated by
textX. For example, if a property \texttt{assignment} of \emph{obj}
is not null, we add the variable name stored in property \texttt{assignment.name}
(on the r.h.s. in the DSL) and ``='' to the generated code. Further
on, if the property \texttt{chainOps} of \emph{obj} is not null, then
we iterate over a list stored in this property. For each list element
we call a method to recursively handle each sub-type of a \emph{ChainOperation}
AST node.

After acquiring some familiarity with the textX framework it becomes
a relatively straightforward tasks to implement new DSL elements.
Also the required (Python) code has reasonable size. In detail, code
generator for our current DSL comprises about 240 LOC (Lines of Code,
including comments) and 3 Python classes.

Different target frameworks are handled as sub-classes of a class
for framework-agnostic code generation. Such sub-classes are typically
small, which ensures that code generation for additional frameworks
can be easily implemented. In our case, a subclass for Pandas code
generation has 55 LOC and 12 short methods, and the subclass for Spark
has 51 LOC and 11 methods.

\input{040-evaluation.tex}

\input{050-related_work.tex}\input{060-conclusions.tex}

\bibliographystyle{ACM-Reference-Format}
\bibliography{zotero-refs-updated-automatically-with-lyz}

\end{document}

%% file: 011-abstract.tex
Data analysis is at the core of scientific studies, a prominent task
that researchers and practitioners typically undertake by programming
their own set of automated scripts. While there is no shortage of
tools and languages available for designing data analysis pipelines,
users spend substantial effort in learning the specifics of such languages/tools
and often design solutions too project specific to be reused in future
studies. Furthermore, users need to put further effort into making
their code scalable, as parallel implementations are typically more
complex.

We address these problems by proposing an advanced code recommendation
tool which facilitates developing data science scripts. Users formulate
their intentions in a human-readable Domain Specific Language (DSL)
for dataframe manipulation and analysis. The DSL statements can be
converted into executable Python code during editing. To avoid the
need to learn the DSL and increase user-friendliness, our tool supports
code completion in mainstream IDEs and editors. Moreover, DSL statements
can generate executable code for different data analysis frameworks
(currently we support Pandas and PySpark). Overall, our approach attempts
to accelerate programming of common data analysis tasks and to facilitate
the conversion of the implementations between frameworks.

In a preliminary assessment based on a popular data processing tutorial,
our tool was able to fully cover 9 out of 14 processing steps for
Pandas and 10 out of 16 for PySpark, while partially covering 4 processing
steps for each of the frameworks.

%% file: 040-evaluation.tex
\section{Preliminary Evaluation} \label{sec:Evaluation}

A proper evaluation of the usefulness of our approach would require
a well-designed user study with a sufficient number of participants.
Due to time constrains, we refrain from performing such analysis and
focus instead on assessing the coverage of our initial DSL implementation.
To that aim, in this preliminary evaluation we answer the following
question: Given a real-usage scenario, what fraction of the analysis
and data processing steps are covered by our current DSL design?

We first need to select a scenario that represents a typical data
analysis task. Due to the popularity and demand for data analysis
in current scientific landscape, tutorials are very common in the
web. For our evaluation, we select a popular DataCamp tutorial ``Apache
Spark Tutorial: ML with PySpark''\footnote{\href{https://www.datacamp.com/community/tutorials/apache-spark-tutorial-machine-learning}{https://www.datacamp.com/community/tutorials/apache-spark-tutorial-machine-learning}}
that provides a high-quality use-case for using PySpark dataframes
for data processing and data analysis. This tutorial has multiple
topics, ranging from Spark installation to the application of conventional
machine-learning methods for Big Data. For this initial assessment,
we focus only on the topics within the scope targeted by our DSL:
data exploration and data pre-processing. The two topics in this tutorial
contain a total of 16 \textit{processing steps}. We define a processing
step loosely as the smallest part of the code that can be executed
on its own (typically single-liners), or in a single block of code
in this tutorial.

To evaluate our DSL in context of Spark, we use unchanged source code
from the tutorial. For Pandas, we manually translate each processing
step into Pandas code, which yields only 14 processing steps, as some
steps (such as creating RDDs to populate dataframes) are specific
to PySpark. We do not use (another) tutorial directly for Pandas as
it was impossible to find a tutorial with similar purpose/scenario
as to the one for Spark. Moreover, many introductory tutorials for
Pandas explain functions related to selecting individual values (e.g.
\texttt{.loc} and \texttt{.iloc}), and functions related to the (row)
index data structure of Pandas. Such functions make Pandas code fundamentally
difficult to translate to other programming paradigms or frameworks
(e.g. SQL, Spark), and typically make the code non-scalable. In our
scenario, we assume that a developer considers the scenario of massive
data sets and will avoid such functions right from the onset. 

In our evaluation we attempt to express a basic processing step with
our DSL. For each such step, we estimate whether it can be expressed
in the DSL completely, whether DSL needs additional (Python) code,
whether we need to substantially change the generated code, or if
our DSL cannot express this step at all. We present in Table~\ref{tab:results_evaluation}
the aggregated results of this initial assessment. Overall, our DSL
fully covers 64.2\% of Pandas processing steps and 62.5\% of PySpark
processing steps in the tutorial. In 4 processing steps for Pandas
and PySpark, the translated code did not completely match the expected
goal and users would have to add Python code (either as new parameters
or function call) to fulfill their initial intention. We find no case
where the generated code needs to be rewritten but three processing
steps (one in Pandas and two in PySpark) could not be expresed or
translated by our DSL (see NS in Table \ref{tab:results_evaluation}). 

\begin{table}
\caption{\label{tab:results_evaluation}Results of our DSL coverage assessment
for Pandas and PySpark code translation. }

\begin{tabular}{lll}
\toprule 
Category & Pandas & PySpark\tabularnewline
\midrule 
Fully translated (FT) & 9/14 (64.2\%) & 10/16 (62.5\%)\tabularnewline
Code added (CA) & 4/14 (28.5\%) & 4/16 (25.0\%)\tabularnewline
Code modified (CM) & 0/14 (0.0\%) & 0/16 (0.0\%)\tabularnewline
DSL not suitable (NS) & 1/14 (7.1\%) & 2/16 (12.5\%)\tabularnewline
\bottomrule
\end{tabular}
\end{table}
In Table~\ref{tab:examples_translation}, we detail the processing
steps our DSL fail to cover in its completion (CA) or could not support
at all (NS). From the 11 partially failed steps, 6 were only covered
by a similarly equivalent function (L1). In this particular case,
our DSL generated the code to retrieve the top N rows in a dataframe
through the \texttt{head} function call, while in the tutorial the
\texttt{show} function is preferred. Both functions are similar in
practice, but \texttt{show} prints the top N rows in the standard
output, while \texttt{head} returns top N rows as a new dataframe.
In another case (L2), when sorting a dataframe, our DSL assumes the
sorting in the ascending order, hence omitting the optional parameter
for descending sorting (\texttt{ascending=False}). The L3 evidences
one limitation of our DSL, the lack of context awareness. In the current
implementation, the DSL does not index or complete code from functions
defined by the programmer and hence should not be used as a code recommender.
The last case (L4), is the support for lambda functions not yet implemented
in our tool. 

\begin{table*}
\caption{\label{tab:examples_translation}Examples of DSL limitation in fully
translating our studied tutorial into PySpark code. The ``Cat''
column references the categories of Table \ref{tab:examples_translation}.
The expected code written in \textbf{\textit{\textcolor{red}{red}}}
and bold is currently not supported by our DSL.}

\small

\begin{tabular}{lllllll}
\toprule 
ID & Cat. & Limitation & DSL code (\texttt{\#\#} is omitted) & Generated code & Expected generated code & \#\tabularnewline
\midrule 
L1 & CA & Target function not supported & on df: return top\_N 10 & df.head(10) & df.\textbf{\textit{\textcolor{red}{show}}}(10) & 6\tabularnewline
L2 & CA & Parameter not supported & on df : sort\_by col & df.sort('col') & df.sort('col', \textbf{\textit{\textcolor{red}{ascending=False}}}) & 2\tabularnewline
L3 & NS & Custom user-functions not supported & -- & -- & df = \textbf{\textit{\textcolor{red}{convertColumn}}}(df, columns,
FloatType()) & 2\tabularnewline
L4 & NS & Lambda functions not supported & -- & -- & df = rdd.\textbf{\textcolor{red}{\emph{apply}}}\textbf{\textcolor{red}{(}}\textbf{\textcolor{red}{\emph{lambda}}}\textbf{\textcolor{red}{{}
}}\textbf{\textcolor{red}{\emph{x}}}\textbf{\textcolor{red}{: }}\textbf{\textcolor{red}{\emph{x}}}\textbf{\textcolor{red}{{}
/ 10).}}\textbf{\textcolor{red}{\emph{toDF}}}\textbf{\textcolor{red}{()}} & 1\tabularnewline
\bottomrule
\end{tabular}
\end{table*}

%% file: 050-related_work.tex
\section{Related Work} \label{sec:Related-Work}

Domains relevant to our work are \emph{low-code data analysis}, \emph{accelerated
scripting and coding}, and \emph{Domain Specific Languages}.

\emph{Low-code data analysis}. Multiple research fields tackle the
challenge of making the process of data analysis and transformation
more user-friendly and accelerating scripting and automation of processing.
The essential directions are: visual analytics \citep{macinnes2010visualclassification},
mixed-initiative systems \citep{horvitz1999principles}, \citep{makonin2016mixedinitiative},
facilitating user involvement in the data analysis or processing activities
\citep{doan2017humanintheloop}, \citep{rahman2018icarusminimizing},
learning data transformations by examples \citep{le2014flashextract},
\citep{smith2016mapreduce}, \citep{jin2017foofahtransforming}, \citep{raza2017automated},
and data wrangling in various flavors \citep{raman2001potters}, \citep{kandel2011wrangler},
\citep{stonebraker2013datacuration}, \citep{heer2015predictive},
\citep{rahman2018icarusminimizing}.

Data wrangling (or data munging) refers to the process of interactive
data transformations on structured or unstructured data. The most
mature tool in this domain is Wrangler \citep{kandel2011wrangler}
which was commercialized by Trifacta \citep{machlis2016datawrangling},
and recently offered by Google as the Google Cloud service \citep{gogle2017clouddataprep}.
Another popular tool is OpenRefine \citep{verborgh2013usingopenrefine}
(originally GoogleRefine) which allows batch processing of tabular
data by menu selection and a Domain Specific Language (DSL) named
GREL. Albeit similar to Wrangler, it has more restricted functionality
and offers no support for very large data sets.

The concept behind these tools is Predictive Interaction \citep{heer2015predictive}.
Its key elements are: (i) real-time preview of effects of code on
a processed data set, (ii) code recommendations based on the context
of user interactions and data, and (iii) a DSL to describe data transformations
in a way easily accessible to users. A disadvantage of the Wrangler
tool is the fact that it is a closed (and commercial) eco-system.
This creates a serious barrier for interoperability with mainstream
libraries or frameworks. Moreover, its DSL has a limited expressiveness
(focusing on data preparation only), and extending this DSL requires
developing User Defined Functions.

Learning data transformations by examples \citep{le2014flashextract},
\citep{smith2016mapreduce}, \citep{jin2017foofahtransforming}, \citep{raza2017automated}
is a special case of the program synthesis techniques. Such approaches
(while still immature) offer a promise to greatly facilitate complex
data analysis, especially for users with no or little programming
skills. The (quite sophisticated) methods here include constraint-based
program synthesis, program sketching \citep{solar-lezama2008program},
version space algebra \citep{le2014flashextract,gulwani2011automating},
or searching in a state space graph \citep{jin2017foofahtransforming}.
In context of data extraction, transformation, and analysis, several
interesting applications have been proposed \citep{gulwani2016programming},
including extracting relations from spreadsheets \citep{barowy2015flashrelate},
data transformations \citep{jin2017foofahtransforming}, or synthesizing
regular expressions. So far, only the FlashFill approach \citep{gulwani2015automating}
has been practically relevant and available to many users (as a component
of Excel 2013 and \emph{ConvertFrom-String} cmdlet in PowerShell).
We consider program synthesis as a possible extension of our work.

\emph{Accelerating scripting and coding} (and as a special case, end-user
development/end-user programming) comprises a multitude of approaches
from software engineering. The most visible progress in this category
stems from novel programming languages, introduction of software processes
such as Scrum, proliferation of software testing, and advances in
development tools (including Intelligent Development Environments,
or syntax/error checkers). Nevertheless, the impact of the individual
measures on programmers\textquoteright{} productivity is hard to measure.
Other noteworthy approaches include program synthesis (discussed above),
visual programming via dataflow languages \citep{johnston2004advances},
block programming languages \citep{bau2017learnable}, and Domain
Specific Languages (DSLs) \citep{fowler2010domainspecific,kosar2016domainspecific}
(discussed below), or more generally Language-Oriented Programming
\citep{felleisen2015theracket}.

In the context of data analysis, dataflow languages \citep{johnston2004advances,gotze2016rewriting}
have gained some popularity via tools such as \citep{2018top21} or
KNIME \citep{berthold2008knimethe}. Such approaches can greatly accelerate
creation of small data processing pipelines and have also proven suitable
for educational purposes. On the other hand, they tend to slow down
more experienced programmers (even smallest operation like \textquotedblleft +\textquotedblright{}
must be selected from menu/palette and connected to other blocks),
do not provide an intuitive support for controls structures, and lack
interoperability with other tools or libraries. For these reasons,
they are rarely used in larger projects.

\emph{Domain Specific Languages} (DSLs) \citep{fowler2010domainspecific},
\citep{kats2010thespoofax}, \citep{dejanovic2017textxa}, have proven
useful in a multitude of medium to large-scale projects by introducing
highly readable and concise code with support for higher-level operations.
While the underlying ``theory'' and scientific interest is still
modest \citep{kats2010thespoofax}, \citep{felleisen2015theracket},
\citep{gupta2015languagebased}, DSLs are becoming increasingly popular
in industry (for example, the industrial-grade database management
system SAP HANA uses internally over 200 DSLs).

A particular flavor of DSLs are internal or embedded DSLs which can
seamlessly inter-operate with the underlying (typically general-purpose)
language. However, internal DSLs offer only limited range of syntax
and are typically not supported by IDEs. Contrary to this, external
DSLs admit almost any syntax, and modern DSL engineering tools (like
MPS \citep{campagne2016themps}, Xtest \citep{bettini2016implementing},
textX \citep{dejanovic2017textxa}, or Spoofax \citep{kats2010thespoofax})
provide ``automatic'' editing support tools (syntax checking and
code completion) for them. The disadvantage of external DSLs is the
difficulty of interaction with (general-purpose) languages. Paired
with this is increased development effort in a scenario where DSL
capabilities are not sufficient, and e.g. writing project-specific
User Defined Functions become necessary.

In our approach we generate code for a general-purpose language from
an external DSL during the editing process, which largely eliminates
the interoperability barrier. We also implemented a special Language
Server to provide coding assistance to both our DSL and the ``embedding''
general-purpose language (here Python).

%% file: 060-conclusions.tex
\section{Conclusions and Future Work} \label{sec:Future-Work-and-Conclusions}

We proposed a DSL-based approach to support data scientists in writing
code for common tasks related to table analysis and processing. We
use external DSL to express such operations in a human-readable form,
and generate executable Python code directly during editing. In this
way we circumvent the frequently encountered problem of insufficient
expressiveness of a DSL, since developers can directly use Python
code to address more special cases (outside the power of the DSL).
Our prototype works with a large number of IDEs and editors (all supporting
the Language Server Protocol), and provides editing support (code
recommendations) for the DSL.

Moreover, users can generate code for Pandas ``data wrangling''
library, or for Apache Spark. This facilitates a transition from low-effort
yet typically non-scalable scripts (in Pandas) suitable for smaller
data sets to highly scalable scripts in Spark. Our preliminary evaluation
shows that for typical data pre-processing tasks, our DSL is capable
of generating complete code in 10 out of 16 cases for Apache Spark,
and in 9 out of 14 cases for Pandas.

Despite of these promising results, there is still a lot of work to
be done to understand and to address the challenges of our approach,
and to provide tools of practical value. Our future work will include
a controlled user study with interviews in order to identify such
challenges, and verify our hypotheses on user behavior. We will also
implement more code generation targets for our DSL, including R (with
dplyr/tidyR) and Matlab. Another option is to provide and evaluate
a DSL for deep learning frameworks like TensorFlow, Microsoft Cognitive
Toolkit, or PyTorch.

Further work related to the editing support will include mechanisms
for synchronizing DSL and the generated code, e.g. by marking DSL
code which is no longer in-sync with the Python code. Another option
here is to add support for ``type providers'' known from .NET languages,
i.e. editor/compiler recommendations for column names and types of
the actual data sets processed in a script.

%% file: 010_main.bbl

\begin{thebibliography}{36}


\ifx \showCODEN    \undefined \def \showCODEN     #1{\unskip}     \fi
\ifx \showDOI      \undefined \def \showDOI       #1{#1}\fi
\ifx \showISBNx    \undefined \def \showISBNx     #1{\unskip}     \fi
\ifx \showISBNxiii \undefined \def \showISBNxiii  #1{\unskip}     \fi
\ifx \showISSN     \undefined \def \showISSN      #1{\unskip}     \fi
\ifx \showLCCN     \undefined \def \showLCCN      #1{\unskip}     \fi
\ifx \shownote     \undefined \def \shownote      #1{#1}          \fi
\ifx \showarticletitle \undefined \def \showarticletitle #1{#1}   \fi
\ifx \showURL      \undefined \def \showURL       {\relax}        \fi
\providecommand\bibfield[2]{#2}
\providecommand\bibinfo[2]{#2}
\providecommand\natexlab[1]{#1}
\providecommand\showeprint[2][]{arXiv:#2}

\bibitem[\protect\citeauthoryear{??}{201}{2018}]%
        {2018top21}
 \bibinfo{year}{2018}\natexlab{}.
\newblock \bibinfo{title}{Top 21 {Self} {Service} {Data} {Preparation}
  {Software} - {Compare} {Reviews}, {Features}, {Pricing} in 2019}.
\newblock
\newblock
\urldef\tempurl%
\url{https://www.predictiveanalyticstoday.com/data-preparation-tools-and-platforms/}
\showURL{%
\tempurl}


\bibitem[\protect\citeauthoryear{Barowy, Gulwani, Hart, and Zorn}{Barowy
  et~al\mbox{.}}{2015}]%
        {barowy2015flashrelate}
\bibfield{author}{\bibinfo{person}{Daniel~W. Barowy}, \bibinfo{person}{Sumit
  Gulwani}, \bibinfo{person}{Ted Hart}, {and} \bibinfo{person}{Benjamin Zorn}.}
  \bibinfo{year}{2015}\natexlab{}.
\newblock \showarticletitle{{FlashRelate}: {Extracting} relational data from
  semi-structured spreadsheets using examples}. In
  \bibinfo{booktitle}{\emph{Proceedings of the 36th {ACM} {SIGPLAN}
  {Conference} on {Programming} {Language} {Design} and {Implementation}}}.
  \bibinfo{publisher}{ACM}, \bibinfo{pages}{218--228}.
\newblock
\urldef\tempurl%
\url{http://dl.acm.org/citation.cfm?id=2737952}
\showURL{%
\tempurl}


\bibitem[\protect\citeauthoryear{Bau, Gray, Kelleher, Sheldon, and Turbak}{Bau
  et~al\mbox{.}}{2017}]%
        {bau2017learnable}
\bibfield{author}{\bibinfo{person}{David Bau}, \bibinfo{person}{Jeff Gray},
  \bibinfo{person}{Caitlin Kelleher}, \bibinfo{person}{Josh Sheldon}, {and}
  \bibinfo{person}{Franklyn Turbak}.} \bibinfo{year}{2017}\natexlab{}.
\newblock \showarticletitle{Learnable {Programming}: {Blocks} and {Beyond}}.
\newblock \bibinfo{journal}{\emph{Commun. ACM}} \bibinfo{volume}{60},
  \bibinfo{number}{6} (\bibinfo{date}{May} \bibinfo{year}{2017}),
  \bibinfo{pages}{72--80}.
\newblock
\showISSN{0001-0782}
\urldef\tempurl%
\url{https://doi.org/10.1145/3015455}
\showDOI{\tempurl}


\bibitem[\protect\citeauthoryear{Berthold, Cebron, Dill, Gabriel, K{\"o}tter,
  Meinl, Ohl, Sieb, Thiel, and Wiswedel}{Berthold et~al\mbox{.}}{2008}]%
        {berthold2008knimethe}
\bibfield{author}{\bibinfo{person}{Michael~R. Berthold},
  \bibinfo{person}{Nicolas Cebron}, \bibinfo{person}{Fabian Dill},
  \bibinfo{person}{Thomas~R. Gabriel}, \bibinfo{person}{Tobias K{\"o}tter},
  \bibinfo{person}{Thorsten Meinl}, \bibinfo{person}{Peter Ohl},
  \bibinfo{person}{Christoph Sieb}, \bibinfo{person}{Kilian Thiel}, {and}
  \bibinfo{person}{Bernd Wiswedel}.} \bibinfo{year}{2008}\natexlab{}.
\newblock \showarticletitle{{KNIME}: {The} {Konstanz} {Information} {Miner}}.
\newblock In \bibinfo{booktitle}{\emph{Data {Analysis}, {Machine} {Learning}
  and {Applications}}}. \bibinfo{publisher}{Springer, Berlin, Heidelberg},
  \bibinfo{pages}{319--326}.
\newblock
\showISBNx{978-3-540-78239-1 978-3-540-78246-9}
\urldef\tempurl%
\url{https://doi.org/10.1007/978-3-540-78246-9_38}
\showDOI{\tempurl}


\bibitem[\protect\citeauthoryear{Bettini}{Bettini}{2016}]%
        {bettini2016implementing}
\bibfield{author}{\bibinfo{person}{Lorenzo Bettini}.}
  \bibinfo{year}{2016}\natexlab{}.
\newblock \bibinfo{booktitle}{\emph{Implementing {Domain} {Specific}
  {Languages} with {Xtext} and {Xtend} - {Second} {Edition}}
  (\bibinfo{edition}{2nd} ed.)}.
\newblock \bibinfo{publisher}{Packt Publishing}.
\newblock
\showISBNx{978-1-78646-496-5}


\bibitem[\protect\citeauthoryear{Bezanson}{Bezanson}{2015}]%
        {bezanson2015abstraction}
\bibfield{author}{\bibinfo{person}{Jeffrey~Werner Bezanson}.}
  \bibinfo{year}{2015}\natexlab{}.
\newblock \emph{\bibinfo{title}{Abstraction in technical computing [{Julia}
  language]}}.
\newblock Thesis. \bibinfo{school}{Massachusetts Institute of Technology}.
\newblock
\urldef\tempurl%
\url{http://dspace.mit.edu/handle/1721.1/99811}
\showURL{%
\tempurl}


\bibitem[\protect\citeauthoryear{Campagne}{Campagne}{2016}]%
        {campagne2016themps}
\bibfield{author}{\bibinfo{person}{Fabien Campagne}.}
  \bibinfo{year}{2016}\natexlab{}.
\newblock \bibinfo{booktitle}{\emph{The {MPS} {Language} {Workbench} {Volume}
  {I}: {The} {Meta} {Programming} {System} ({Volume} 1)}
  (\bibinfo{edition}{3rd} ed.)}.
\newblock \bibinfo{publisher}{CreateSpace Independent Publishing Platform},
  \bibinfo{address}{USA}.
\newblock
\showISBNx{978-1-5305-3335-0}


\bibitem[\protect\citeauthoryear{Corp.}{Corp.}{2019}]%
        {microsoftcorp.2019language}
\bibfield{author}{\bibinfo{person}{Microsoft Corp.}}
  \bibinfo{year}{2019}\natexlab{}.
\newblock \bibinfo{title}{Language {Server} {Protocol} {Specification}}.
\newblock
\newblock
\urldef\tempurl%
\url{https://microsoft.github.io/language-server-protocol/specification}
\showURL{%
\tempurl}


\bibitem[\protect\citeauthoryear{Dejanovi{\'c}, Vaderna, Milosavljevi{\'c}, and
  Vukovi{\'c}}{Dejanovi{\'c} et~al\mbox{.}}{2017}]%
        {dejanovic2017textxa}
\bibfield{author}{\bibinfo{person}{I. Dejanovi{\'c}}, \bibinfo{person}{R.
  Vaderna}, \bibinfo{person}{G. Milosavljevi{\'c}}, {and} \bibinfo{person}{{\v
  Z}. Vukovi{\'c}}.} \bibinfo{year}{2017}\natexlab{}.
\newblock \showarticletitle{{TextX}: {A} {Python} tool for {Domain}-{Specific}
  {Languages} implementation}.
\newblock \bibinfo{journal}{\emph{Knowledge-Based Systems}}
  \bibinfo{volume}{115} (\bibinfo{date}{Jan.} \bibinfo{year}{2017}),
  \bibinfo{pages}{1--4}.
\newblock
\showISSN{0950-7051}
\urldef\tempurl%
\url{https://doi.org/10.1016/j.knosys.2016.10.023}
\showDOI{\tempurl}


\bibitem[\protect\citeauthoryear{Doan, Ardalan, Ballard, Das, Govind, Konda,
  Li, Mudgal, Paulson, Suganthan, and Zhang}{Doan et~al\mbox{.}}{2017}]%
        {doan2017humanintheloop}
\bibfield{author}{\bibinfo{person}{AnHai Doan}, \bibinfo{person}{Adel Ardalan},
  \bibinfo{person}{Jeffrey Ballard}, \bibinfo{person}{Sanjib Das},
  \bibinfo{person}{Yash Govind}, \bibinfo{person}{Pradap Konda},
  \bibinfo{person}{Han Li}, \bibinfo{person}{Sidharth Mudgal},
  \bibinfo{person}{Erik Paulson}, \bibinfo{person}{G.~C.~Paul Suganthan}, {and}
  \bibinfo{person}{Haojun Zhang}.} \bibinfo{year}{2017}\natexlab{}.
\newblock \showarticletitle{Human-in-the-{Loop} {Challenges} for {Entity}
  {Matching}: {A} {Midterm} {Report}}. In \bibinfo{booktitle}{\emph{Proceedings
  of the 2Nd {Workshop} on {Human}-{In}-the-{Loop} {Data} {Analytics}}}
  \emph{(\bibinfo{series}{{HILDA}'17})}. \bibinfo{publisher}{ACM},
  \bibinfo{address}{New York, NY, USA}, \bibinfo{pages}{12:1--12:6}.
\newblock
\showISBNx{978-1-4503-5029-7}
\urldef\tempurl%
\url{https://doi.org/10.1145/3077257.3077268}
\showDOI{\tempurl}


\bibitem[\protect\citeauthoryear{Felleisen, Findler, Flatt, Krishnamurthi,
  Barzilay, McCarthy, Tobin-Hochstadt, and Herbstritt}{Felleisen
  et~al\mbox{.}}{2015}]%
        {felleisen2015theracket}
\bibfield{author}{\bibinfo{person}{Matthias Felleisen},
  \bibinfo{person}{Robert~Bruce Findler}, \bibinfo{person}{Matthew Flatt},
  \bibinfo{person}{Shriram Krishnamurthi}, \bibinfo{person}{Eli Barzilay},
  \bibinfo{person}{Jay McCarthy}, \bibinfo{person}{Sam Tobin-Hochstadt}, {and}
  \bibinfo{person}{Marc Herbstritt}.} \bibinfo{year}{2015}\natexlab{}.
\newblock \bibinfo{booktitle}{\emph{The {Racket} {Manifesto}}}.
\newblock \bibinfo{type}{{T}echnical {R}eport}. \bibinfo{institution}{Schloss
  Dagstuhl - Leibniz-Zentrum fuer Informatik GmbH, Wadern/Saarbruecken,
  Germany}. \bibinfo{pages}{--} pages.
\newblock
\urldef\tempurl%
\url{https://doi.org/10.4230/LIPIcs.SNAPL.2015.113}
\showDOI{\tempurl}


\bibitem[\protect\citeauthoryear{Fowler}{Fowler}{2010}]%
        {fowler2010domainspecific}
\bibfield{author}{\bibinfo{person}{Martin Fowler}.}
  \bibinfo{year}{2010}\natexlab{}.
\newblock \bibinfo{booktitle}{\emph{Domain {Specific} {Languages}}
  (\bibinfo{edition}{1st} ed.)}.
\newblock \bibinfo{publisher}{Addison-Wesley Professional}.
\newblock
\showISBNx{0-321-71294-3 978-0-321-71294-3}
\newblock
\shownote{00681.}


\bibitem[\protect\citeauthoryear{Gogle}{Gogle}{2017}]%
        {gogle2017clouddataprep}
\bibfield{author}{\bibinfo{person}{Gogle}.} \bibinfo{year}{2017}\natexlab{}.
\newblock \bibinfo{title}{Cloud {Dataprep} - {Data} {Preparation} and {Data}
  {Cleansing}}.
\newblock
\newblock
\urldef\tempurl%
\url{https://cloud.google.com/dataprep/}
\showURL{%
\tempurl}


\bibitem[\protect\citeauthoryear{G{\"o}tze and Sattler}{G{\"o}tze and
  Sattler}{2016}]%
        {gotze2016rewriting}
\bibfield{author}{\bibinfo{person}{Philipp G{\"o}tze} {and}
  \bibinfo{person}{Kai-Uwe Sattler}.} \bibinfo{year}{2016}\natexlab{}.
\newblock \showarticletitle{Rewriting and {Code} {Generation} for {Dataflow}
  {Programs}}.
\newblock \bibinfo{journal}{\emph{GvD}} (\bibinfo{year}{2016}),
  \bibinfo{pages}{6}.
\newblock


\bibitem[\protect\citeauthoryear{Gulwani}{Gulwani}{2011}]%
        {gulwani2011automating}
\bibfield{author}{\bibinfo{person}{Sumit Gulwani}.}
  \bibinfo{year}{2011}\natexlab{}.
\newblock \showarticletitle{Automating string processing in spreadsheets using
  input-output examples}. In \bibinfo{booktitle}{\emph{{ACM} {SIGPLAN}
  {Notices}}}, Vol.~\bibinfo{volume}{46}. \bibinfo{publisher}{ACM},
  \bibinfo{pages}{317--330}.
\newblock
\urldef\tempurl%
\url{http://dl.acm.org/citation.cfm?id=1926423}
\showURL{%
\tempurl}


\bibitem[\protect\citeauthoryear{Gulwani}{Gulwani}{2015}]%
        {gulwani2015automating}
\bibfield{author}{\bibinfo{person}{Sumit Gulwani}.}
  \bibinfo{year}{2015}\natexlab{}.
\newblock \showarticletitle{Automating {Repetitive} {Tasks} for the {Masses}}.
  In \bibinfo{booktitle}{\emph{Proceedings of the 42Nd {Annual} {ACM}
  {SIGPLAN}-{SIGACT} {Symposium} on {Principles} of {Programming} {Languages}}}
  \emph{(\bibinfo{series}{{POPL} '15})}. \bibinfo{publisher}{ACM},
  \bibinfo{address}{New York, NY, USA}, \bibinfo{pages}{1--2}.
\newblock
\showISBNx{978-1-4503-3300-9}
\urldef\tempurl%
\url{https://doi.org/10.1145/2676726.2682621}
\showDOI{\tempurl}


\bibitem[\protect\citeauthoryear{Gulwani}{Gulwani}{2016}]%
        {gulwani2016programming}
\bibfield{author}{\bibinfo{person}{Sumit Gulwani}.}
  \bibinfo{year}{2016}\natexlab{}.
\newblock \showarticletitle{Programming by {Examples} (and its {Applications}
  in {Data} {Wrangling})}. In \bibinfo{booktitle}{\emph{Verification and
  {Synthesis} of {Correct} and {Secure} {Systems}}}. \bibinfo{publisher}{IOS
  Press}.
\newblock
\urldef\tempurl%
\url{https://www.microsoft.com/en-us/research/publication/programming-examples-applications-data-wrangling/}
\showURL{%
\tempurl}


\bibitem[\protect\citeauthoryear{Gupta}{Gupta}{2015}]%
        {gupta2015languagebased}
\bibfield{author}{\bibinfo{person}{Gopal Gupta}.}
  \bibinfo{year}{2015}\natexlab{}.
\newblock \showarticletitle{Language-based {Software} {Engineering}}.
\newblock \bibinfo{journal}{\emph{Sci. Comput. Program.}} \bibinfo{volume}{97},
  \bibinfo{number}{P1} (\bibinfo{date}{Jan.} \bibinfo{year}{2015}),
  \bibinfo{pages}{37--40}.
\newblock
\showISSN{0167-6423}
\urldef\tempurl%
\url{https://doi.org/10.1016/j.scico.2014.02.010}
\showDOI{\tempurl}


\bibitem[\protect\citeauthoryear{Heer, Hellerstein, and Kandel}{Heer
  et~al\mbox{.}}{2015}]%
        {heer2015predictive}
\bibfield{author}{\bibinfo{person}{Jeffrey Heer}, \bibinfo{person}{Joseph
  Hellerstein}, {and} \bibinfo{person}{Sean Kandel}.}
  \bibinfo{year}{2015}\natexlab{}.
\newblock \showarticletitle{Predictive {Interaction} for {Data}
  {Transformation}}. In \bibinfo{booktitle}{\emph{Conference on {Innovative}
  {Data} {Systems} {Research} ({CIDR})}}.
\newblock
\urldef\tempurl%
\url{http://idl.cs.washington.edu/papers/predictive-interaction}
\showURL{%
\tempurl}


\bibitem[\protect\citeauthoryear{Horvitz}{Horvitz}{1999}]%
        {horvitz1999principles}
\bibfield{author}{\bibinfo{person}{Eric Horvitz}.}
  \bibinfo{year}{1999}\natexlab{}.
\newblock \showarticletitle{Principles of {Mixed}-initiative {User}
  {Interfaces}}. In \bibinfo{booktitle}{\emph{Proceedings of the {SIGCHI}
  {Conference} on {Human} {Factors} in {Computing} {Systems}}}
  \emph{(\bibinfo{series}{{CHI} '99})}. \bibinfo{publisher}{ACM},
  \bibinfo{address}{New York, NY, USA}, \bibinfo{pages}{159--166}.
\newblock
\showISBNx{978-0-201-48559-2}
\urldef\tempurl%
\url{https://doi.org/10.1145/302979.303030}
\showDOI{\tempurl}


\bibitem[\protect\citeauthoryear{Jin, Anderson, Cafarella, and Jagadish}{Jin
  et~al\mbox{.}}{2017}]%
        {jin2017foofahtransforming}
\bibfield{author}{\bibinfo{person}{Zhongjun Jin}, \bibinfo{person}{Michael~R.
  Anderson}, \bibinfo{person}{Michael Cafarella}, {and} \bibinfo{person}{H.~V.
  Jagadish}.} \bibinfo{year}{2017}\natexlab{}.
\newblock \showarticletitle{Foofah: {Transforming} {Data} {By} {Example}}.
  \bibinfo{publisher}{ACM Press}, \bibinfo{pages}{683--698}.
\newblock
\showISBNx{978-1-4503-4197-4}
\urldef\tempurl%
\url{https://doi.org/10.1145/3035918.3064034}
\showDOI{\tempurl}


\bibitem[\protect\citeauthoryear{Johnston, Hanna, and Millar}{Johnston
  et~al\mbox{.}}{2004}]%
        {johnston2004advances}
\bibfield{author}{\bibinfo{person}{Wesley~M. Johnston},
  \bibinfo{person}{J.~R.~Paul Hanna}, {and} \bibinfo{person}{Richard~J.
  Millar}.} \bibinfo{year}{2004}\natexlab{}.
\newblock \showarticletitle{Advances in {Dataflow} {Programming} {Languages}}.
\newblock \bibinfo{journal}{\emph{ACM Comput. Surv.}} \bibinfo{volume}{36},
  \bibinfo{number}{1} (\bibinfo{date}{March} \bibinfo{year}{2004}),
  \bibinfo{pages}{1--34}.
\newblock
\showISSN{0360-0300}
\urldef\tempurl%
\url{https://doi.org/10.1145/1013208.1013209}
\showDOI{\tempurl}


\bibitem[\protect\citeauthoryear{Kandel, Paepcke, Hellerstein, and Heer}{Kandel
  et~al\mbox{.}}{2011}]%
        {kandel2011wrangler}
\bibfield{author}{\bibinfo{person}{Sean Kandel}, \bibinfo{person}{Andreas
  Paepcke}, \bibinfo{person}{Joseph Hellerstein}, {and}
  \bibinfo{person}{Jeffrey Heer}.} \bibinfo{year}{2011}\natexlab{}.
\newblock \showarticletitle{Wrangler: {Interactive} {Visual} {Specification} of
  {Data} {Transformation} {Scripts}}. In \bibinfo{booktitle}{\emph{Proceedings
  of the {SIGCHI} {Conference} on {Human} {Factors} in {Computing} {Systems}}}
  \emph{(\bibinfo{series}{{CHI} '11})}. \bibinfo{publisher}{ACM},
  \bibinfo{address}{New York, NY, USA}, \bibinfo{pages}{3363--3372}.
\newblock
\showISBNx{978-1-4503-0228-9}
\urldef\tempurl%
\url{https://doi.org/10.1145/1978942.1979444}
\showDOI{\tempurl}


\bibitem[\protect\citeauthoryear{Kats and Visser}{Kats and Visser}{2010}]%
        {kats2010thespoofax}
\bibfield{author}{\bibinfo{person}{Lennart~C.L. Kats} {and}
  \bibinfo{person}{Eelco Visser}.} \bibinfo{year}{2010}\natexlab{}.
\newblock \showarticletitle{The {Spoofax} {Language} {Workbench}: {Rules} for
  {Declarative} {Specification} of {Languages} and {IDEs}}. In
  \bibinfo{booktitle}{\emph{Proceedings of the {ACM} {International}
  {Conference} on {Object} {Oriented} {Programming} {Systems} {Languages} and
  {Applications}}} \emph{(\bibinfo{series}{{OOPSLA} '10})}.
  \bibinfo{publisher}{ACM}, \bibinfo{address}{New York, NY, USA},
  \bibinfo{pages}{444--463}.
\newblock
\showISBNx{978-1-4503-0203-6}
\urldef\tempurl%
\url{https://doi.org/10.1145/1869459.1869497}
\showDOI{\tempurl}
\newblock
\shownote{event-place: Reno/Tahoe, Nevada, USA.}


\bibitem[\protect\citeauthoryear{Kosar, Bohra, and Mernik}{Kosar
  et~al\mbox{.}}{2016}]%
        {kosar2016domainspecific}
\bibfield{author}{\bibinfo{person}{Toma{\v z} Kosar}, \bibinfo{person}{Sudev
  Bohra}, {and} \bibinfo{person}{Marjan Mernik}.}
  \bibinfo{year}{2016}\natexlab{}.
\newblock \showarticletitle{Domain-{Specific} {Languages}: {A} {Systematic}
  {Mapping} {Study}}.
\newblock \bibinfo{journal}{\emph{Information and Software Technology}}
  \bibinfo{volume}{71} (\bibinfo{date}{March} \bibinfo{year}{2016}),
  \bibinfo{pages}{77--91}.
\newblock
\showISSN{0950-5849}
\urldef\tempurl%
\url{https://doi.org/10.1016/j.infsof.2015.11.001}
\showDOI{\tempurl}


\bibitem[\protect\citeauthoryear{Le and Gulwani}{Le and Gulwani}{2014}]%
        {le2014flashextract}
\bibfield{author}{\bibinfo{person}{Vu Le} {and} \bibinfo{person}{Sumit
  Gulwani}.} \bibinfo{year}{2014}\natexlab{}.
\newblock \showarticletitle{{FlashExtract}: {A} {Framework} for {Data}
  {Extraction} by {Examples}}. In \bibinfo{booktitle}{\emph{Proceedings of the
  35th {ACM} {SIGPLAN} {Conference} on {Programming} {Language} {Design} and
  {Implementation}}} \emph{(\bibinfo{series}{{PLDI} '14})}.
  \bibinfo{publisher}{ACM}, \bibinfo{address}{New York, NY, USA},
  \bibinfo{pages}{542--553}.
\newblock
\showISBNx{978-1-4503-2784-8}
\urldef\tempurl%
\url{https://doi.org/10.1145/2594291.2594333}
\showDOI{\tempurl}


\bibitem[\protect\citeauthoryear{Machlis}{Machlis}{2016}]%
        {machlis2016datawrangling}
\bibfield{author}{\bibinfo{person}{Sharon Machlis}.}
  \bibinfo{year}{2016}\natexlab{}.
\newblock \bibinfo{title}{Data wrangling tool {Trifacta} aims to ease analysis
  pain}.
\newblock
\newblock
\urldef\tempurl%
\url{http://www.computerworld.com/article/3104769/data-analytics/data-wrangling-tool-trifacta-aims-to-ease-analysis-pain.html}
\showURL{%
\tempurl}


\bibitem[\protect\citeauthoryear{MacInnes, Santosa, and Wright}{MacInnes
  et~al\mbox{.}}{2010}]%
        {macinnes2010visualclassification}
\bibfield{author}{\bibinfo{person}{Joseph MacInnes}, \bibinfo{person}{Stephanie
  Santosa}, {and} \bibinfo{person}{William Wright}.}
  \bibinfo{year}{2010}\natexlab{}.
\newblock \showarticletitle{Visual {Classification}: {Expert} {Knowledge}
  {Guides} {Machine} {Learning}}.
\newblock \bibinfo{journal}{\emph{IEEE Comput. Graph. Appl.}}
  \bibinfo{volume}{30}, \bibinfo{number}{1} (\bibinfo{date}{Jan.}
  \bibinfo{year}{2010}), \bibinfo{pages}{8--14}.
\newblock
\showISSN{0272-1716}
\urldef\tempurl%
\url{https://doi.org/10.1109/MCG.2010.18}
\showDOI{\tempurl}


\bibitem[\protect\citeauthoryear{Makonin, McVeigh, Stuerzlinger, Tran, and
  Popowich}{Makonin et~al\mbox{.}}{2016}]%
        {makonin2016mixedinitiative}
\bibfield{author}{\bibinfo{person}{Stephen Makonin}, \bibinfo{person}{Daniel
  McVeigh}, \bibinfo{person}{Wolfgang Stuerzlinger}, \bibinfo{person}{Khoa
  Tran}, {and} \bibinfo{person}{Fred Popowich}.}
  \bibinfo{year}{2016}\natexlab{}.
\newblock \showarticletitle{Mixed-{Initiative} for {Big} {Data}: {The}
  {Intersection} of {Human} + {Visual} {Analytics} + {Prediction}}.
  \bibinfo{publisher}{IEEE}, \bibinfo{pages}{1427--1436}.
\newblock
\showISBNx{978-0-7695-5670-3}
\urldef\tempurl%
\url{https://doi.org/10.1109/HICSS.2016.181}
\showDOI{\tempurl}


\bibitem[\protect\citeauthoryear{Rahman, Hebert, and Nandi}{Rahman
  et~al\mbox{.}}{2018}]%
        {rahman2018icarusminimizing}
\bibfield{author}{\bibinfo{person}{Protiva Rahman}, \bibinfo{person}{Courtney
  Hebert}, {and} \bibinfo{person}{Arnab Nandi}.}
  \bibinfo{year}{2018}\natexlab{}.
\newblock \showarticletitle{{ICARUS}: {Minimizing} {Human} {Effort} in
  {Iterative} {Data} {Completion}}.
\newblock \bibinfo{journal}{\emph{Proc. VLDB Endow.}} \bibinfo{volume}{11},
  \bibinfo{number}{13} (\bibinfo{date}{Sept.} \bibinfo{year}{2018}),
  \bibinfo{pages}{2263--2276}.
\newblock
\showISSN{2150-8097}
\urldef\tempurl%
\url{https://doi.org/10.14778/3275366.3284970}
\showDOI{\tempurl}


\bibitem[\protect\citeauthoryear{Raman and Hellerstein}{Raman and
  Hellerstein}{2001}]%
        {raman2001potters}
\bibfield{author}{\bibinfo{person}{Vijayshankar Raman} {and}
  \bibinfo{person}{Joseph~M. Hellerstein}.} \bibinfo{year}{2001}\natexlab{}.
\newblock \showarticletitle{Potter's {Wheel}: {An} {Interactive} {Data}
  {Cleaning} {System}}. In \bibinfo{booktitle}{\emph{Proceedings of the 27th
  {International} {Conference} on {Very} {Large} {Data} {Bases}}}
  \emph{(\bibinfo{series}{{VLDB} '01})}. \bibinfo{publisher}{Morgan Kaufmann
  Publishers Inc.}, \bibinfo{address}{San Francisco, CA, USA},
  \bibinfo{pages}{381--390}.
\newblock
\showISBNx{1-55860-804-4}
\urldef\tempurl%
\url{http://dl.acm.org/citation.cfm?id=645927.672045}
\showURL{%
\tempurl}
\newblock
\shownote{00411.}


\bibitem[\protect\citeauthoryear{Raza and Gulwani}{Raza and Gulwani}{2017}]%
        {raza2017automated}
\bibfield{author}{\bibinfo{person}{Mohammad Raza} {and} \bibinfo{person}{Sumit
  Gulwani}.} \bibinfo{year}{2017}\natexlab{}.
\newblock \showarticletitle{Automated {Data} {Extraction} {Using} {Predictive}
  {Program} {Synthesis}}. In \bibinfo{booktitle}{\emph{Proceedings of the
  {Thirty}-{First} {AAAI} {Conference} on {Artificial} {Intelligence},
  {February} 4-9, 2017, {San} {Francisco}, {California}, {USA}.}},
  \bibfield{editor}{\bibinfo{person}{Satinder~P. Singh} {and}
  \bibinfo{person}{Shaul Markovitch}} (Eds.). \bibinfo{publisher}{AAAI Press},
  \bibinfo{pages}{882--890}.
\newblock
\urldef\tempurl%
\url{http://aaai.org/ocs/index.php/AAAI/AAAI17/paper/view/15034}
\showURL{%
\tempurl}


\bibitem[\protect\citeauthoryear{Smith and Albarghouthi}{Smith and
  Albarghouthi}{2016}]%
        {smith2016mapreduce}
\bibfield{author}{\bibinfo{person}{Calvin Smith} {and} \bibinfo{person}{Aws
  Albarghouthi}.} \bibinfo{year}{2016}\natexlab{}.
\newblock \showarticletitle{{MapReduce} {Program} {Synthesis}}. In
  \bibinfo{booktitle}{\emph{Proceedings of the 37th {ACM} {SIGPLAN}
  {Conference} on {Programming} {Language} {Design} and {Implementation}}}
  \emph{(\bibinfo{series}{{PLDI} '16})}. \bibinfo{publisher}{ACM},
  \bibinfo{address}{New York, NY, USA}, \bibinfo{pages}{326--340}.
\newblock
\showISBNx{978-1-4503-4261-2}
\urldef\tempurl%
\url{https://doi.org/10.1145/2908080.2908102}
\showDOI{\tempurl}


\bibitem[\protect\citeauthoryear{Solar-Lezama}{Solar-Lezama}{2008}]%
        {solar-lezama2008program}
\bibfield{author}{\bibinfo{person}{Armando Solar-Lezama}.}
  \bibinfo{year}{2008}\natexlab{}.
\newblock \emph{\bibinfo{title}{Program {Synthesis} by {Sketching}}}.
\newblock {PhD} {Thesis}. \bibinfo{school}{University of California at
  Berkeley}, \bibinfo{address}{Berkeley, CA, USA}.
\newblock


\bibitem[\protect\citeauthoryear{Stonebraker, Ilyas, Zdonik, Beskales, and
  Pagan}{Stonebraker et~al\mbox{.}}{2013}]%
        {stonebraker2013datacuration}
\bibfield{author}{\bibinfo{person}{Michael Stonebraker},
  \bibinfo{person}{Ihab~F Ilyas}, \bibinfo{person}{Stan Zdonik},
  \bibinfo{person}{George Beskales}, {and} \bibinfo{person}{Alexander Pagan}.}
  \bibinfo{year}{2013}\natexlab{}.
\newblock \showarticletitle{Data {Curation} at {Scale}: {The} {Data} {Tamer}
  {System}}.
\newblock \bibinfo{journal}{\emph{6th Biennial Conference on Innovative Data
  Systems Research}} (\bibinfo{year}{2013}).
\newblock


\bibitem[\protect\citeauthoryear{Verborgh and De~Wilde}{Verborgh and
  De~Wilde}{2013}]%
        {verborgh2013usingopenrefine}
\bibfield{author}{\bibinfo{person}{Ruben Verborgh} {and} \bibinfo{person}{Max
  De~Wilde}.} \bibinfo{year}{2013}\natexlab{}.
\newblock \bibinfo{booktitle}{\emph{Using {OpenRefine}} (\bibinfo{edition}{1st
  new edition edition} ed.)}.
\newblock \bibinfo{publisher}{Packt Publishing}.
\newblock
\urldef\tempurl%
\url{http://openrefine.org/}
\showURL{%
\tempurl}


\end{thebibliography}
